\documentclass[showpacs,twocolumn,prb,aps,color,superscriptaddress]{revtex4}
\usepackage{graphicx}

\begin{document}
 
\hsize\textwidth\columnwidth\hsize\csname @twocolumnfalse\endcsname

\title{Transition from band insulator to Mott insulator in one dimension:\\
Critical behavior and phase diagram}
\author{Jizhong Lou}
\affiliation{Institute of Theoretical Physics, P. O. Box 2735, Beijing 100080, China}
\affiliation{Department of Physics, University of Nevada, Las Vegas, Nevada 89154}
\author{Shaojin Qin}
\affiliation{Institute of Theoretical Physics, P. O. Box 2735, Beijing 100080, China}
\affiliation{Department of Physics, Kyushu University, Hakozaki, Higashi-ku, Fukuoka 812-8581, 
Japan} 
\author{Tao Xiang}
\affiliation{Institute of Theoretical Physics, P. O. Box 2735, Beijing 100080, China}
\author{Changfeng Chen}
\affiliation{Department of Physics, University of Nevada, Las Vegas, Nevada 89154}
\author{Guang-Shan Tian}
\affiliation{Department of Physics, Peking University, Beijing 100087, China}
\author{Zhaobin Su}
\affiliation{Institute of Theoretical Physics, P. O. Box 2735, Beijing 100080, China}
\date{\today}

\begin{abstract}
We report a systematic study of the transition from band insulator (BI) to Mott 
insulator (MI) in a one-dimensional Hubbard model with an on-site Coulomb interaction
$U$ and an alternating periodic site potential $V$.  We employ both the zero-temperature 
density matrix renormalization group (DMRG) method to determine the gap and critical 
behavior of the system and the finite-temperature transfer matrix renormalization 
group (TMRG) method to evaluate the thermodynamic properties.  We find two critical 
points at $U = U_c$ and $U = U_s$ that separate the BI and MI phases for a given $V$.
A charge-neutral spin-singlet exciton band develops in the BI phase ($U<U_c$)
and drops below the band gap when $U$ exceeds a special point $U_e$.  The exciton
gap closes at the first critical point $U_c$ while the charge and spin gaps persist
and coincide between $U_c<U<U_s$ where the system is dimerized.  Both the charge and spin 
gaps collapse at $U = U_s$ when the transition to the MI phase occurs.  In the MI 
phase ($U>U_s$) the charge gap increases almost linearly with $U$ while the spin gap
remains zero.  These findings clarify earlier published results on the same model
and offer new insights into several important issues regarding appropriate scaling 
analysis of DMRG data and a full physical picture for the delicate nature of the phase
transitions driven by electron correlation.  The present work provides a comprehensive 
understanding for the critical behavior and phase diagram for the transition from BI to 
MI in one-dimensional correlated electron systems with a periodic alternating site potential.
\end{abstract}

\draft
\maketitle

\section{Introduction}

The nature of the insulating ground state of interacting electron systems has 
been a subject of long-standing interest and debate in condensed matter physics. 
Because of strong quantum effects caused by spatial confinement and technical
advantages for theoretical treatment, one-dimensional (1D) electron systems have 
been most extensively studied\cite{soly,hald}.  Strong correlation effects in 1D
lead to the separation (decoupling) of charge and spin degrees of freedom.  
Starting from a gapless phase with charge-spin separated excitations, 
interactions can drive the system into new phases of different characteristics 
with (i) gapful charge excitations only, (ii) gapful spin excitations only, or 
(iii) co-existing gapful charge and spin excitations.  For phases with both charge and 
spin excitations gapful, the charge and spin degrees of freedom are rarely decoupled.  
Despite these findings, there remain important unresolved issues regarding 
the quantum nature of the insulating state in 1D interacting electron systems and the
phase transitions driven by the electron correlation.  The first issue concerns the 
establishment of an accurate phase diagram and the critical behavior near the phase 
boundaries.  Secondly, a band insulator (BI) with quasi-particle excitations typically 
cannot be characterized by charge and spin excitations.  A proper characterization scheme 
needs to be developed.  Most importantly, the nature of the correlation-driven transition 
from BI to Mott insulator (MI) is still not fully understood.

There has been considerable recent interest in the study of a prototype 
one-dimensional model for ferroelectric perovskites for the understanding of the 
response of strongly correlated electron systems with lattice distortions. These 
efforts have raised and addressed some fundamental issues in the nature of the quantum 
phase transition and the related critical behavior in 1D interacting electron systems.  
Earlier works \cite{Egami,Nagaosa,Ishihara} mainly deal with the effects of strong 
electron correlation on the electron-lattice interaction and the polarization effects 
in the insulator.  Quantum phase transitions and the characterization of the insulating 
state are the focus of more recent work.\cite{Ort,Schonhammer,Resta,tsu99,naka,tsuc,wilk,anus,yone,capr,tori} In particular, an issue of fascinating 
debate is the nature of the transition (or crossover) from the BI phase to the MI phases.  
Gidopoulos {\it et al.} showed \cite{Gidopoulos} that due to the reversal of 
inversion symmetry of the ground state from BI to MI, there is a critical point 
for spin excitations.  However, for charge excitations the critical behavior is 
less clear.  Recently, Fabrizio {\it et al.} \cite{Fabrizio} developed an 
effective field theory for this problem and showed that there are two continuous 
transitions from BI to MI .  One is a spin transition of the Kosterlitz-Thouless type 
at a critical point $U=U_s$, and the other is an exciton transition at an Ising 
critical point $U=U_c<U_s$ where the exciton gap closes.  Between $U_c$ and $U_s$, 
the site-parity is spontaneously broken and the system is characterized by a doubly 
degenerate, dimerized ground state.  These results raise interesting questions about the
structure of the ground-state phase diagram of 1D interacting electron systems and
the characterization of the critical behavior near the transition points from the
BI phase to the MI phase.

The model Hamiltonian for the system of interest is defined in the Hubbard formalism 
at half-filling \cite{Ishihara,Ort,Schonhammer,tsu99,avig} 
\begin{eqnarray}
H &=&	\sum_{i\sigma }\left[ 
	-t\left( c_{i\sigma }^{\dagger }c_{i+1\sigma }+
	{\rm h.c.}\right) +V(-1)^in_{i\sigma }
	\right]  \nonumber \\
  & &	+U\sum_i\left( n_{i\uparrow }-\frac 12\right) 
	\left( n_{i\downarrow }- \frac 12\right) ,  
\label{Ham}
\end{eqnarray}
where $c_{i\sigma }^{\dagger }$ and $n_{i\sigma }$ are the electron creation and 
number operators at site $i$, $U>0$ is the on-site Coulomb repulsion, and $V$ is the 
staggered site chemical potential.   This model captures the key ingredients of 
one-dimensional correlated insulators with mixed ionic-covalent characters, such as 
oxide dielectric materials\cite{rest} and quasi-1D organic charge-transfer 
complexes.\cite{torr} It incorporates covalency, ionicity, and strong electron 
correlation.\cite{Egami,Nagaosa} 

In this paper, we present the results of extensive calculations for Hamiltonian (1) 
using both the zero-temperature density matrix renormalization group (DMRG) 
method \cite{White92} and the finite-temperature transfer matrix renormalization 
group (TMRG) method \cite{Bursill,Wang,Shibata}.  These methods have been demonstrated 
to be highly accurate for 1D interacting electron systems.  Our aim is to systematically 
examine and clarify issues raised in recent work and to provide a comprehensive 
understanding for the transition from BI to MI in one dimension.  We show detailed results 
on the gap and critical behavior.  We find that a charge-neutral spin-singlet exciton 
band forms in the BI phase and 
drops below the band gap as $U$ increases beyond a special point $U_e$.  With 
increasing $U$ the excitons then condense and the system enters a dimerized phase, 
followed by the closure of the quasi-particle (both spin and charge) gap when the 
system enters the MI phase.  We clarify basic concepts on charge and spin excitations 
studied in recent work.  Our results support the conclusion of 
Fabrizio {\it et al.}\cite{Fabrizio} on the existence of two critical points for the 
transition from BI to MI.  We also present detailed results on the formation of the 
exciton band and the scaling behavior near the critical points.  In addition, we
carry out TMRG calculations to study thermodynamic properties to further elucidate the 
gap and critical behavior of the system.

\section{Low-energy excitations}
\label{elem}

To properly characterize the BI and MI phases and establish the phase diagram, we need to
evaluate the behavior of several low-energy excitations, including the charge and spin 
excitations and an exciton excitation that will be used to characterize the BI phase.
We calculate the following three excitation gaps defined on a finite 1D lattice of length 
$L$ (chosen as an even integer) at half-filling: (i) the singlet exciton gap $\Delta_e(L)$, 
(ii) the charge gap $\Delta _c(L)$, and (iii) the spin-triplet gap $\Delta _s(L)$, 
\begin{eqnarray}
\Delta _e(L) &=&
	E_1(\frac L2,\frac L2)-E_0(\frac L2,\frac L2),  
	\label{gape} \\
\Delta _c(L) &=&
	E_0(\frac L2+1,\frac L2)+E_0(\frac L2-1,\frac L2)-
	2E_0(\frac L2,\frac L2),  \label{gapc} \\
\Delta _s(L) &=&E_0(\frac L2+1,\frac L2-1)-E_0(\frac L2,\frac L2),
\label{gaps}
\end{eqnarray}
where $E_0(N_{\uparrow },N_{\downarrow })$ is the lowest energy of the system with 
$N_{\uparrow }$ up and $N_{\downarrow }$ down spin electrons, and $E_1(N_{\uparrow },
N_{\downarrow })$ is the lowest energy of the singlet excitations.  For the charge gap 
$\Delta_c$, there is an alternative definition 
\begin{equation}
\Delta _c(L)=E(\frac L2+1,\frac L2+1)-E(\frac L2,\frac L2). 
\label{gapcc}
\end{equation}
For the model studied in this work, these two definitions on the charge gap give the same 
results in the thermodynamic limit.  This is supported by our numerical calculations.
However, since the numerical accuracy is much higher in calculating $\Delta_c$ by using
Eq. (\ref{gapc}) than by using Eq. (\ref{gapcc}), we use Eq. (\ref{gapc}) in all the
reported calculations.

Although $\Delta_c$ and $\Delta_s$ are usually considered to be the charge gap and the spin 
gap, respectively, in the literature,  $\Delta_c$ in fact is the chemical potential jump for 
particles in the system.  It measures the chemical potential jump of putting a particle into 
or taking a particle out of the system.  Only when the charge and spin excitations are 
separated and the spin gap is zero, $\Delta_c$ is equal to the charge excitation gap as, 
for example, in the standard Hubbard model (Eq. (\ref{Ham}) with $V=0$ or $U\to\infty$).  
When the charge and spin excitations are not separated, or when the spin gap is not zero, 
the chemical potential jump is not equal to the charge excitation gap.

At half-filling, the first excitation state can be either a spin singlet or a spin triplet
and $\Delta_e$ never exceeds $\Delta_s$.  When $\Delta_e < \Delta_s$, the first excitation
state must be a charge-neutral spin-singlet state; otherwise, $\Delta_e$ equals $\Delta_s$ 
and measures the excitation gap of an exciton band.

In the presence of a nonzero $V$ in the Hamiltonian, the MI phase is reached when $U$ is 
large (including the limit $U\to\infty$).  In the MI phase, it is well understood that the 
charge gap is non-zero, but the spin gap is zero.  Therefore $\Delta_c$ is the charge 
excitation gap.  Meanwhile, both $\Delta_s$ and $\Delta_e$ are zero and the gapless 
elementary excitations in the system are spinons.\cite{ess} 

In the BI phase, all the elementary excitations are gapful.  Let $u^\dag_{k\sigma}$ and 
$d^\dag_{k\sigma}$  be the creation operators for particles in the upper conducting band 
and for the holes in the lower valence band.  The Hamiltonian for free particles 
at $U=0$ is
\begin{equation}
H = \sum_{k\sigma} \varepsilon_k 
	\left( u^\dag_{k\sigma} u_{k\sigma} - 
	v^\dag_{k\sigma} v_{k\sigma} \right),
\label{freeham}
\end{equation}
where $k$ is the momentum, $\sigma$ is the spin index, and 
$\varepsilon _k=\sqrt{V^2+4t^2\cos ^2k}$.
We have $u_{k\sigma}|GS\rangle=v^\dag_{k\sigma}|GS\rangle=0$ for the ground state 
$|GS\rangle$ at half filling with $u_{k\sigma}$ being the annihilation operator for 
electrons in the conduction band, and $v^\dag_{k\sigma}$ the creation operator for
electrons in the valence band.  For particle-hole excitations with one particle and one hole
in the system, it is clear that $\Delta_c=\Delta_s=\Delta_e=2V$.

When $U$ is small in the BI phase, the particle or hole excitations become dressed 
quasi-particles, and $\Delta_c$ measures the chemical potential jump of the particles.  
Since the charge and spin degrees of freedom are not separated,  $\Delta_c$ is not exactly 
the "charge gap" derived from the gapless charge-spin separated excitations in the field 
theoretical approach.  If the particles and holes are not bounded, then 
$\Delta_c=\Delta_s=\Delta_e$.  However, the particle-hole excitations may bound to form 
excitons, and result in an exciton gap smaller than  $\Delta_c$.  It is confirmed by our 
DMRG calculations shown below  that $\Delta_e$ equals $\Delta_c$ when $U$ is small, but 
becomes smaller than $\Delta_c$ when $U$ exceeds a special point $U_e$, indicating the 
formation of singlet excitons.  Meanwhile, the calculations show that $\Delta_s = \Delta_c$ 
is always observed, indicating that there is no triplet exciton formation in the BI phase.  

The DMRG calculations also show that in the BI phase the exciton gap and quasi-particle 
excitation gap decrease and approach zero with increasing $U$.  The exciton gap closes 
at the first critical point $U_c$, and the system enters a phase where the excitons condense 
to form dimerized ground state.  At the second critical point $U_s$ the quasi-particle gap 
also closes ($\Delta_e =\Delta_s = \Delta_c=0 $).  When $U$ exceeds $U_s$, $\Delta_e$ and
$\Delta_s$ remain zero but $\Delta_c$ increases almost linearly with $U$.  
Therefore, $U_s$ is the point where the quasi-particle excitation gap in the BI phase 
collapses, and spinons in the MI phase form.  This picture is consistent with that of the 
recent field theoretical studies\cite{Fabrizio}.

From the physical picture outlined above, it is clear that there are three special points 
$U_e$, $U_c$, and $U_s$ along the $U$ scale.  Among them 
$U_c$ and $U_s$ are two critical points separating the BI and MI phases while $U_e$ is
a special point signaling the formation of the spin-singlet exciton band in the BI phase.
In the thermodynamic limit for a given $V$ the system can be divided into the following
regions:  \\
\begin{enumerate}
\item $0<U<U_{e}$, $\Delta _e = \Delta_c = \Delta_s \ne 0$, 
	gapful quasi-particle excitations only.
\item $U_{e}<U<U_{c}$, $ 0< \Delta _e < \Delta_c = \Delta_s$, 
	gapful quasi-particle excitations coexist with singlet particle-hole 
	excitations that bound to form excitons.
\item $U=U_{c}$, $ \Delta _e = 0$, $\Delta_c = \Delta_s >0$, 
	the system is critical for exciton excitations.
\item $U_{c}<U<U_{s}$, $\Delta _e < \Delta_c = \Delta_s$,
	excitons condense, and the system is dimerized.
\item $U=U_{s}$, $\Delta _e = \Delta_c = \Delta_s=0$, 
	the system is critical for quasi-particle excitations.
\item $U>U_{s}$, $\Delta _e = \Delta_s = 0$, $\Delta_c>0$, 
	gapless spinon excitations in the MI phase;  the particle-hole 
	picture breaks down.
\end{enumerate}

In the following we present the calculated results on the gap and critical behavior
leading to the establishment of the phase diagram.  In all reported calculations, 
free boundary conditions are used in the zero-temperature DMRG calculation. For a 
finite size system, we can show rigorously using the variational principle that the 
lowest energy state of the Hamiltonian (\ref {Ham}) with free boundary conditions in 
each $(N_{\uparrow },N_{\downarrow }) $ subspace is non-degenerate except for an up-down
spin degeneracy\cite{Xiang}. Therefore there is no level crossing with the lowest 
energy state in the $(N_{\uparrow },N_{\downarrow }) $ subspace and 
$E_0(N_{\uparrow },N_{\downarrow })$ is an analytic function of $U$ and $V$. From this 
property and Eqs. (\ref{gapc}) and (\ref {gaps}), we can further show that $\Delta _c(L)$ and 
$\Delta _s(L)$ are also analytic functions of $U$ and $V$.  In this work we focus on two 
values of the staggered potential $V$=0.3t and 1.0t and study the behavior of the three 
gaps introduced above in response to the on-site Coulomb repulsion $U$.

\section{Excitation Gaps}
\label{energygap}

For fermion systems, the truncation error of DMRG iterations is generally much 
smaller than that of spin systems when the same number of optimal states are retained. 
The efficiency of the finite system DMRG method is related to the truncation error;
the bigger the truncation error is, the larger the improvement of the finite lattice 
sweeping can make.  We used both finite and infinite lattice DMRG algorithms in testing 
calculations.  We find that the improvement of the ground-state energy made with the finite 
lattice sweeping is very small when a large number of states are retained.  A better way 
to increase the accuracy of the results is using the infinity lattice approach by retaining 
more states. 

\begin{figure}[h!]
\includegraphics[width=8.0cm,angle=0]{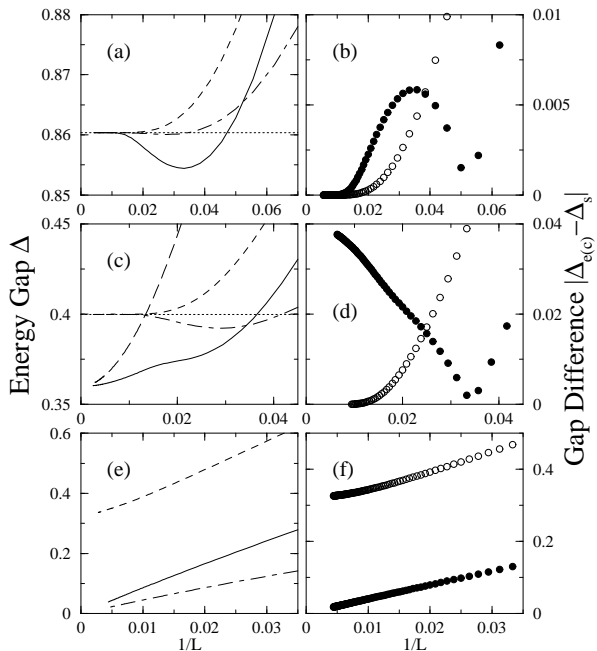}
\caption{Energy gaps vs $1/L$ for $V=1.0t$ and (a) $U=2.0 t$, (c) $U=3.0 t$,
(e) $U=4.0 t$. Solid lines are $\Delta_e$, dot-dashed lines are $\Delta_s$, and short-dashed 
lines are $\Delta_c$. The long-dashed line in (c) represents the gap of the second state in 
the singlet sector $\Delta_c^{(2)}$.  The dotted lines in (a) and (c) denote the thermodynamic 
limit of the corresponding gaps.  Panels (b), (d), and (f) show the absolute values of the 
difference of the charge and exciton gap with the spin gap, where empty circles show 
$\Delta_c - \Delta_s$ and filled circles show $\Delta_e - \Delta_s$.  500 states are retained 
in the DMRG calculations.}
\label{BDG}
\end{figure}

Figure \ref{BDG} shows the behaviors of $\Delta _s$, $\Delta _c$ and $\Delta_e$ as a 
function of $1/L$ for different $U$ at $V=1.0t$. For $V=0.3t$, similar results can be drawn
but the $V=1.0t$ case is more accurate because most of the features can be seen when the 
chain length is short, while in the $V=0.3t$ case, very long chains need to be used to 
obtain the same results.  Figure \ref{BDG} (a) presents the results for $U=2.0 t$ and
clearly shows that the three gaps converge to the same finite value in the thermodynamic limit. 
The difference between these gaps shown in Fig. \ref{BDG} (b) also displays this feature 
clearly.  When the chain length is short, the exciton gap is larger than the spin gap, and 
level crossing happens at a finite chain length where the exciton gap drops lower thereafter. 
The exciton gap decreases continually and reaches a minimum when the chain length increases
further; it then starts to increase and converge to the value of the spin and charge gaps.
For $U=3.0t$, in Fig. \ref{BDG} (c) and (d), the spin gap and charge gap still converge to the 
same value in the thermodynamic limit, but the exciton gap goes to a different value lower
than the other two gaps. In short chains, the exciton gap is still larger than the spin gap, 
but after they cross each other, the exciton gap decreases monotonically. One can also see that
the second state in the singlet sector also crosses the spin and charge gap and converges to 
the lowest state when $L \to \infty$.  In our calculations, we also see that more states in 
the singlet sector cross the spin and charge gap with increasing chain length. This 
shows that the whole spectrum of the exciton sector decreases in value and the exciton gap 
is indeed different from the other two gaps.  For the case of $U=4.0t$, shown in 
Fig. \ref{BDG} (e) and (f), no level crossing for different chain length is detected.
All three gaps decrease monotonically when the chain length increases.  The exciton gap 
and the spin gap approach zero at infinite chain length, indicating that there is no gap
for the exciton and spin sectors. Meanwhile the charge gap approaches a finite value in 
the thermodynamic limit.  

\begin{figure}[h!]
\includegraphics[width=8.0cm,angle=0]{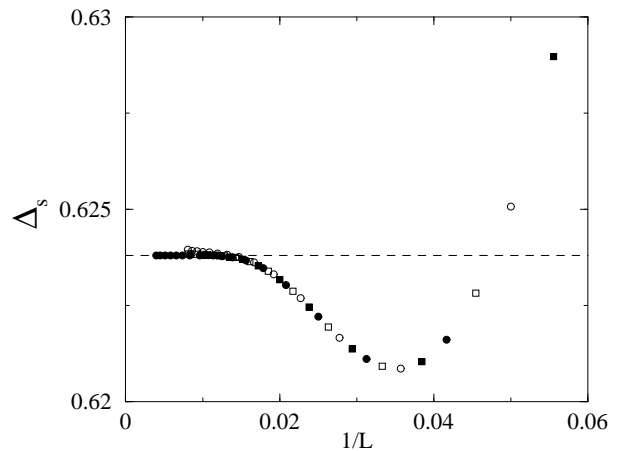}
\caption{The spin gap $\Delta_s$ vs $1/L$ for $V=1.0t$ and $U=2.50t$. The
number of retained states are m=300 (empty circles),400 (empty squares),
500 (filled circles), and 800 (filled squares).}
\label{gaplength}
\end{figure}

For all the cases we have studied, $\Delta _c$ decreases monotonically with increasing $L$. 
However, the size dependence of $\Delta _s$ is more complicated.  In certain ranges of $U/t$ 
and $V/t$ close to the critical regimes, including the case shown in Fig. \ref{BDG} (a) 
and (c), $\Delta _s$ and $\Delta_e$ vary non-monotonically and their minima are located at 
a finite $L=L_{\min }$ rather than at $L=\infty$.  In a recent work \cite{Brune}, the authors 
studied the same model Hamiltonian (\ref{Ham}) using the DMRG method but did not observe such 
non-monotonically behavior, and suggested that such a behavior may be due to the loss of 
accuracy in DMRG calculations when the chain length is increased or due to some intrinsic 
length scale for the spin degree of freedom.  We have carefully examined this issue by
carrying out extensive scaling analysis.  We demonstrate that the non-monotonical behavior 
is not due to the lack of accuracy of the calculations, instead the behavior is a true feature
of Hamiltonian (\ref{Ham}) with the open boundary condition (OBC).  Figure \ref{gaplength}
shows the chain length dependence of the spin gap for $U=2.5 t$ and $V=1.0t$ calculated by 
retaining different numbers of optimal states $m=300, 400, 500$, and $800$.  One can see that 
the minimum occurs at $L \sim 30$; at this length the accuracy of the DMRG calculations are
still very high.  More significantly, the results for different $m$ fall onto 
the same curve (except for the cases of $L > 100$ and  $m=300$).  It shows unambiguously
the existence of the minimum of $\Delta_s$ in its dependence on the chain length.
For the exciton gap $\Delta_e$, the situation is the same. In fact the occurrence of a
gap minimum at a finite $L$ is not an uncommon feature for a system with incommensurate 
low-lying excitations.  It suggests that the spin excitations of the model 
Hamiltonian (\ref{Ham}) maybe incommensurate with a characteristic wave vector defined 
by $2\pi /L_{\min }$ (or $\pi -2\pi /L_{\min }$) in some area of the phase space. 

\begin{figure}[h!]
\includegraphics[width=8.0cm,angle=0]{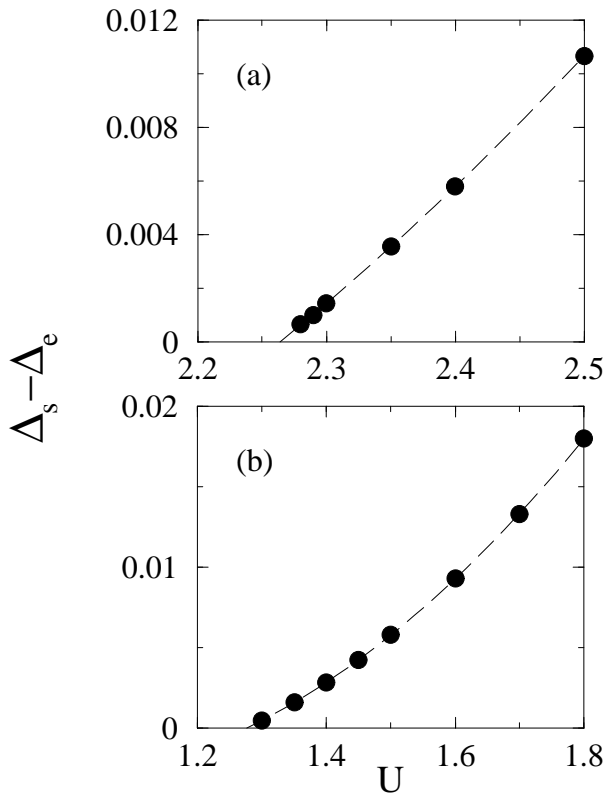}
\caption{The difference between the spin gap and the exciton gap near $U_e$.
(a) V=1.0t, $U_e \sim 2.264 t$.  The dashed fitting line is $0.026(U/t-0.771)(U/t-2.264)$;
(b) V=0.3t, $U_e \sim 1.276 t$.  The dashed fitting line is $0.0282(U/t-0.584)(U/t-1.276)$.}
\label{exc}
\end{figure}

Comparing Fig. \ref{BDG} (a) and (c), it is clear that there is a special point 
$U_e$, where the exciton gap begins to deviate from the spin gap (for $V=1.0t$, 
$2.0t < U_e < 3.0 t$). In both cases, all three excitations are gapful. The system is
in the same (BI) phase as the $U =0$ case, where $\Delta_s=\Delta_c=\Delta_e= 2V$ in
the thermodynamic limit.  Here particle-hole excitations bound into excitons by the 
Coulomb interaction at $U>U_e$.  Fig. \ref{exc} shows the difference $\Delta_s - \Delta_e$
for $V=1.0t$ and $0.3t$.  The fitting to DMRG results gives the critical value $U_e=2.264 t$ 
for $V=1.0t$ and $U_e=1.276t$ for $V=0.3t$.

For finite $L$ we find that $\Delta _c$ is always larger than $\Delta _s$.  In the MI phase, 
$\Delta _c$ is finite but $\Delta _s$ approaches zero in the thermodynamic limit.
In the BI phase, $\Delta _s\ $and $\Delta _c$ always approach the same value 
in the thermodynamic limit. This can be seen either from the asymptotic behaviors of 
$\Delta _s$ and $\Delta _c$ in the limit $L\rightarrow \infty $ [Fig. \ref{BDG} (a) and
(c)] or from the $1/L$ dependence of the difference $\Delta _c-\Delta _s$ 
(Fig. \ref{BDG} (b) and (d)).  For all the cases we have studied, we find that 
$\Delta _c-\Delta _s$ drops monotonically and approaches zero in the limit $1/L\rightarrow 0$ 
even when $\Delta _s$ changes non-monotonically.  For a given $V$, this result holds from 
$U=0$ up to a critical regime where both $\Delta _c$ and $\Delta _s$ 
become smaller than truncation errors. It suggests that $\Delta _c$ and $\Delta _s$ are 
equal in the thermodynamic limit in the entire BI phase.  

When $\Delta _s$ changes 
non-monotonically with $L$, the extrapolation for the spin gap in the limit 
$1/L\rightarrow 0$ becomes subtle. If the data with $L<L_{\min }$ are used in the 
extrapolation, the extrapolated value of $ \Delta _s$ will certainly be smaller than 
the true value. However, if the data with $L>L_{\min }$ is used but $L$ is still not large 
enough to reach the regime where $\Delta _s$ begins to saturate, the extrapolated value of 
$ \Delta _s$ will be larger than the true value (this seems to be the case in 
Ref. \onlinecite{Takada}). For the data shown in Fig. \ref {BDG}(a) and (c), 
these two kinds of extrapolations result in $\Delta _c>$ $\Delta _s$
and $\Delta _c<$ $\Delta _s$, respectively. Both are incorrect. To correctly extrapolate 
$\Delta _s$ in the limit $1/L\rightarrow 0$, data with $L$ much larger than $L_{\min }$
must be used.

For the $U=4.0 t$ case shown in Fig. \ref{BDG} (e), $\Delta _c$ is finite 
but $\Delta _s$ becomes zero in the thermodynamic limit.  This indicates that the system 
is in the same phase as that for $U \to \infty$,  namely the MI phase.  In the MI phase, the 
chain length dependence of the three gaps are monotonical; in addition, there is no 
crossing between $\Delta_s$ and $\Delta_e$ when the chain length varies.  In the 
$L \to \infty$ limit the spin gap $\Delta_s$ is zero in the MI phase suggesting that 
there is a critical point that separates the MI phase from the BI phase.
At this critical point $U_s$, the spin gap vanishes. Because the charge gap is
equal to the spin gap in the BI phase, the charge gap will also vanish at the same 
point $U_s$.  However, when $U$ increases further, the charge gap increases with $U$
while the spin gap remains zero.  Considering that the exciton gap is lower than the spin
gap in the BI phase at $U > U_e$, $\Delta_e$ may vanish before the spin gap and charge gap 
do.  In that case, there should be another critical point $U_c$ signaling the collapse of
the exciton gap.

\begin{figure}[h!]
\includegraphics[width=8.0cm,angle=0]{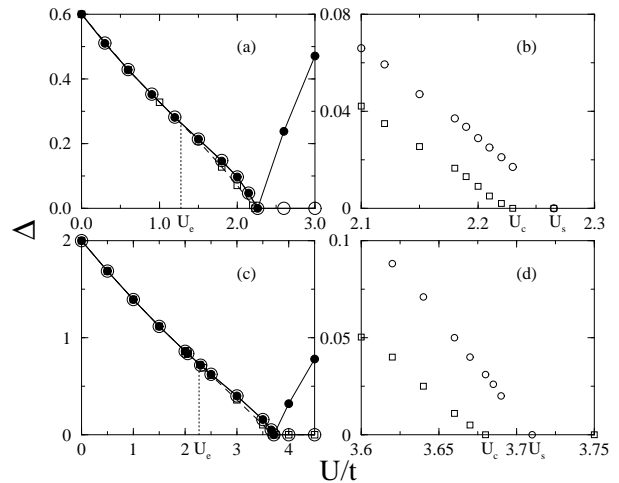}
\caption{The spin (empty circles), charge (solid circles) and exciton (empty squares) gaps
as a function of $U$ for $V =0.3t$ (a) and $V=1.0t$ (c).  Panels (b) and (d) show the $U$ 
dependence of the spin and exciton gaps in the vicinity of the critical region for (a) 
and (c) respectively.  At $U_e$, the exciton gap deviates from the other two gaps and it
collapses first at $U_c \sim 2.225t$ for $V=0.3t$ and $\sim 3.675t$ for $V=1.0t$. The charge 
and spin gaps collapse at $U_s \sim 2.265t$ for $V=0.3t$ and $\sim 3.71t$ for $V=1.0t$. 
For $0<U<U_s$ the spin and charge gaps have the same value in the thermodynamic limit. 
When $U>U_s$, the spin and exciton gap are zero while the charge gap is finite.}
\label{bdu}
\end{figure}

In Fig. \ref{bdu}, we show the $U$ dependence of the three gaps for $V=0.3t$ 
[(a) and (b)] and $V=1.0t$ [(c) and (d)]. For both cases, there are indeed two critical 
points $U_c$ and $U_s$ although they are very close. When $U<U_e$, 
$\Delta_e = \Delta_s = \Delta_c$, and the three gaps decrease almost linearly
with increasing $U$. At $U_e$, the exciton gap splits off and drops below the other two gaps.
At the critical point $U_c$, the exciton gap collapses while the spin and charge gaps still
coincide and remain finite until the second critical point $U_s$ where they both collapse.
At $U > U_s$, the charge gap increases with increasing $U$ while the spin gap and the
exciton gap remain zero in the thermodynamic limit.  Although the accuracy of our
DMRG calculations do not allow a direct assessment of the behavior of the exciton gap for
$U_c < U < U_s$, we believe that $\Delta_e$ is finite in this region (see more detailed 
discussion on this point in the following section).  The extrapolation of the gap behavior 
leads to $U_c \sim 2.225 t$ and $U_s \sim 2.265 t$ for $V=0.3 t$, while $U_c \sim 3.675 t$, 
$U_s \sim 3.71 t$ for $V=1.0 t$.

It is clear that the $U$ dependence of the three gaps is similar for $V=0.3t$ and $V=1.0t$.
It is expected that the same picture is valid for all $V$. Furthermore, $U_e$, $U_c$, and 
$U_s$ all approach the same point $U_{\infty}=2V$ in the $V \rightarrow \infty$ limit.

\section{Critical behavior}
\label{critbeh}

\begin{figure}[h!]
\includegraphics[width=8.0cm,angle=0]{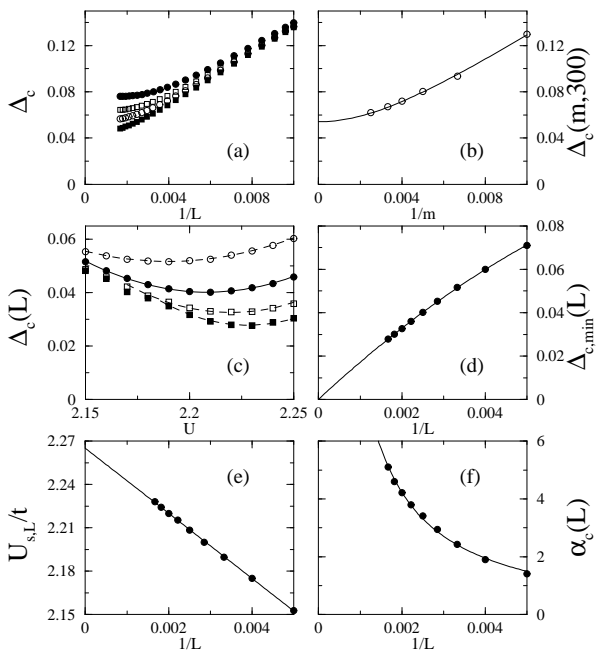}
\caption{The behavior of the charge gap in the vicinity of $U_s$ for $V=0.3t$.  (a) The 
chain length dependence of the charge gap at $U=2.2t$ for different numbers of optimal 
states retained: $m$=200 (filled circles), 250 (empty squares), 300 (empty circles) and 400 
(filled squares); (b) The dependence of the charge gap on the number of the retained states 
at L=300 with $U=2.2t$; (c) The charge gap for $m \to \infty$ with different chain lengths:
$L=$ 300 (empty circles), 400 (filled circles), 500 (empty squares) and 600 (filled circles)
in the vicinity of the critical region; the fitting lines using eq. (\ref{gapc2}) are also 
shown; (d) The chain length dependence of $\Delta_{c,min} (L)$; (e) The chain length 
dependence of $U_{s,L}/t$; (f) The chain length dependence of $\alpha_{c,L}$.}
\label{cgcritical}
\end{figure}

From the gap behavior presented in the previous section, it is clear that there are two 
critical points $U_c$ and $U_s$ for Hamiltonian (\ref{Ham}) for a given $V$.  It is 
important to study the detailed critical behavior near the critical points for the 
understanding of the nature of the BI-to-MI transition. In the following we study the critical
behavior of the system by examining (i) the evolution of the gap behavior near the critical 
points and (ii) the behavior of the ground-state energy of the system. 

\subsection{Analysis of the gap behavior}

The $U$ dependence of the gaps shows that the charge instability occurs at the same point 
as that for the spin transition.  To determine the critical points for the charge and spin 
excitations, we examine when $\Delta _c$ and $\Delta _s $ become zero in the 
limit $L\rightarrow \infty $.  Since the numerical errors are larger than the magnitude of
$\Delta _c$ or $\Delta _s$ in the vicinity of the critical points, it is difficult to 
determine accurately the critical behavior simply from the values of the energy gaps. 
To resolve this issue, we analyze the scaling behavior of $\Delta_c(L)$ around its minimum
with respect to $U$.  However, the DMRG results of $\Delta_c(L)$ depend the number of 
states $m$ retained during the iterations.  When the chain length is long enough, the 
difference due to retaining different number of states show clearly. 
In Fig. \ref{cgcritical} (a), we show the chain length dependence of the charge gap 
at $U=2.22t$ and $V=0.3t$ by keeping different $m$. The difference
is obvious. This problem can be solved by employing the extrapolation in the limit of 
$m\rightarrow \infty $.  For given $U$ and $V$, the charge gap at chain length $L$ and
by keeping $m$ states is $\Delta_c (m, L)$.  By extrapolating to the infinite $m$ limit, 
more accurate result of the charge gap at chain length $L$ can be obtained, 
\begin{equation}
\Delta_c(\infty, L) = \lim_{m \rightarrow \infty} \Delta_c (m,L).
\end{equation}
In Fig. \ref{cgcritical} (b), we show the $1/m$ dependence and the extrapolation procedure 
of the $\Delta_c (m,L)$  for L=300 at $U=2.22t$ and $V=0.3t$. The extrapolated result 
$\Delta_c (\infty, L)$ is considered the exact charge gap $\Delta_c (L)$ at chain length $L$.

The extrapolated charge gap $\Delta_c(\infty, L)$ is a function of $U$ and chain length 
$L$ at a given $V$. By applying the same procedure shown in Fig. \ref{cgcritical} (b), we 
obtain $\Delta_c(\infty, L)$ for different values of $U$ near the critical point $U_s$ for 
a serial of selected $L$.  In Fig. \ref{cgcritical} (c), we show the $U$ dependence of 
$\Delta_c(L)$ ($\Delta_c(\infty, L)$) in the vicinity of the critical point $U_s$ at chain 
length $L=$ 300, 400, 500, and 600, respectively, for $V$=0.3t.  A gap minimum at finite 
chain length is clearly seen.  Assuming $\Delta _{c,\min }(L)$ to be the minimum of 
$\Delta _c(L)$ located at $U_{s,L}$, then around this minimum we can expand 
$\Delta _c(L)$ to the leading order of the parameter $u=U-U_{s,L}$ as 
\begin{equation}
\Delta _c(L)=\Delta _{c,\min }(L)+\alpha _c\left( L\right) u^2+O\left(
u^3\right) .  \label{gapc2}
\end{equation}
Since $\Delta _c(L)$ is an analytic function of $U$, both $\Delta _{c,\min }(L)$ and 
$\alpha _c\left( L\right) $ should be finite. The critical behavior of the charge 
excitations is determined by the properties of $ \Delta _{c,\min }(L)$ and 
$\alpha _c\left( L\right) $ in the limit $ L\rightarrow \infty $. 
If $\Delta _{c,\min }(L)\rightarrow 0$ in the limit $ L\rightarrow \infty $, the charge 
excitation is critical at $ U_s=U_{s,\infty }$, which would be consistent with the discussion 
in the previous section. However, if $\Delta _{c,\min }$ remains finite 
in the limit $L\rightarrow \infty $,  then there is no critical point
for charge excitations and the ground state is insulating in the entire parameter space.

Figure \ref{cgcritical} (d) shows the calculated $\Delta _{c,\min }(L)$ as a function of $1/L$.
The solid curve is a least-square fit of the data and given by 
$\Delta _{c,\min }(L)\approx 17.894/L-729.785/L^2$. Within numerical 
errors, we find that $ \Delta _{c,\min }(L)$ is indeed $0$ in the limit $1/L\rightarrow 0$. 
Figure \ref{cgcritical} (e) shows the $L$ dependence of $U_{s,L}$.  It changes almost
linearly with $1/L$.  Within numerical errors, we find that the data of $U_{s,L}$ are well 
fitted by $U_{s,L}/t \approx 2.265-22.532/L$. Thus the critical point is at $U_c /t =2.265t$,
in full agreement with the value obtained by fitting the gap directly in the previous section.

We now turn to the critical behavior of $\alpha _c\left( L\right) $.  
Figure \ref{cgcritical} (f) shows the $1/L$ 
dependence of $\alpha _c\left( L\right) $ for the case $V=0.3t$. 
The fitting curve (solid line) is given by $ -0.323+0.0091L$. 
The divergence of $\alpha _c\left( L\right) $ suggests that
the derivative of $\Delta _c(L)$ is singular at $U_s$ and 
the leading term in $\Delta _c$ in the thermodynamic limit 
is linear rather than quadratical in $u$, i.e., $\Delta _c(U)\sim \left| U-U_s\right| $. 

\subsection{Analysis of the ground-state energy}
\label{gseng}

The ground-state energy as the zero-temperature free energy can also provide evidence 
for the critical behavior.  However, singularities in the ground-state energy are of higher 
order derivatives with respect to the model parameter for continuous phase transitions.  
As a result, evidence for critical behavior derived from the ground-state energy 
is not as strong as that from the gap behavior, this despite the higher accuracy of the
ground-state energy than that for the gap in the DMRG calculations.

We have calculated the ground-state energy in the critical region by retaining
$m$=800 states and up to chain length $L$=1000.  For Hamiltonian (\ref{Ham}), with open 
boundary conditions, the ground-state energy per site $e_0(L)$ satisfies
\begin{equation}
e_0(L) = \frac{E_0(L)}{L} = \epsilon_0 + \frac{e_b}{L} + \frac{c}{L^2}
+ O(\frac{1}{L^3}),
\end{equation}
here $E_0(L)$ is the ground-state energy for a chain of length $L$, $\epsilon_0$ the 
ground-state energy per site for $L \to \infty$, $e_b$ is the boundary energy 
(surface energy) due to the free boundary condition, and $c=v \pi$ where $v$ is the spin 
wave velocity.  When $U > U_c$, the system is gapful and $c$ should approach zero when 
the chain length is much larger than the correlation length $\xi$.

\begin{figure}[h!]
\includegraphics[width=8.0cm,angle=0]{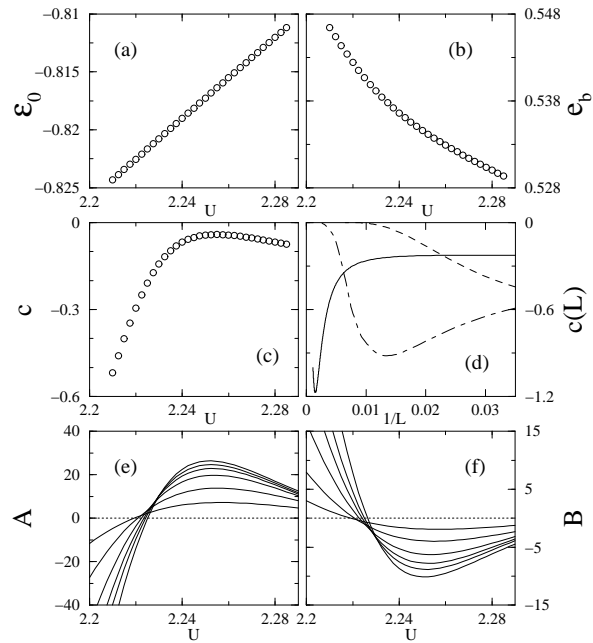}
\caption{(a), (b), (c): The $U$ dependence of $\epsilon_0$, $e_b$ and $c$ obtained from 
fitting the $L>200$ data of the ground-state energy; (d) The chain length dependence of 
c(L) for $U=1.5t$ (dashed line), $U=2.0t$ (dot-dashed line), and $U=2.2t$ (solid line);
(e) $A = 10^6 (\epsilon_0(L) - \epsilon_0(100))$ for L=120, 150, 200, 250, 300, and 400 
(from bottom to top on the $A > 0$ side); (f) $B = 10^3 (e_b(L) - e_b(100))$ for the same 
chain lengths as in (e) (from top to bottom on the $B < 0$ side).}
\label{gse}
\end{figure}

The $V=0.3 t$, $L> 200$ ground-state energy results are fitted directly by 
\begin{equation}
e_0(L) = \epsilon_0 + e_b/ L + c/L^2, 
\label{gsfit}
\end{equation}
and the obtained results are shown in Fig. \ref{gse} (a), (b) and (c).
$\epsilon_0$ and $e_b$ are analytic functions of $U$. 
For $c$, Fig. \ref{gse}(c) shows that it is not only non-zero for $U < U_c$
but also has a fairly large value. This is due to the finite chain length effect.
The value of $c$ depends very sensitively on the chain length range used for fitting.

To analyze the chain length dependence of $c$,  we fit the ground-state energy
for $L-20$, $L$, and $L+20$ using Eq. (\ref{gsfit}) and vary $L$ from 24 to 980. Each 
fitting gives the exact result of $\epsilon_0(L)$, $e_b(L)$ and $c(L)$.  In 
Fig. \ref{gse}(d), we show the $L$ dependence of $c(L)$ for $U=1.5 t$, $U=2.0t$
and $U=2.2t$. It is clear that for $U=1.5t$ and $2.0t$, $c(L)$ becomes zero when 
$L \rightarrow \infty$.  The result of $U=2.0t$ shows a minimum at $L \sim 90$, and 
the $c(L)$ begins to approach zero after the minimum.  For $U=2.2t$, a minimum is also
clearly seen. A comparison of these results leads to the conclusion that for $L \rightarrow
\infty$, $c(L)$ will approach zero. The largest chain length used in the calculation
is not long enough to obtain correct $c(L)$ values. However, the obtained $\epsilon_0(L)$ and
$e_b(L)$ may display the critical behavior of $U_c$ and $U_s$. In Fig. \ref{gse} 
(e) and (f), we show the results of $\epsilon_0 (L) -\epsilon_0(L_0)$ and 
$e_b(L) - e_b(L_0)$ for $L_0=100$.  These results show the existence of the 
critical point $U_c$. $L_0$ can be viewed as a characteristic length of the critical region. 

From the $U$ dependence of $\epsilon_0$, it is possible to examine the type of transition 
at the critical points.  A problem is that the fitting results shown in Fig. \ref{gse}(a) 
include extra errors induced by the fitting method. To avoid this, we analyze the $U$ 
dependence of the ground-state energy using a different approach.

\begin{figure}[h!]
\includegraphics[width=8.0cm,angle=0]{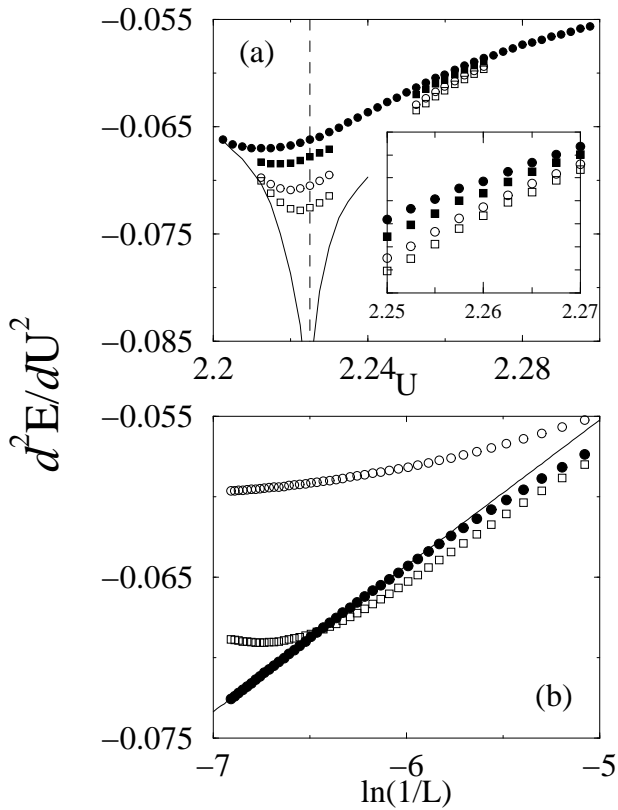}
\caption{(a) ${\it d}^2 e_0/{\it d} U^2$ versus $U$ for $L=500$ (filled circles),
600 (filled squares), 800 (empty circles), 1000 (empty squares).  The solid line is an 
extrapolation of the data to the limit $L \to \infty$. Inset: Enlarged figure in the 
vicinity of the critical point $U_s$.  (b) The dependence of the ${\it d}^2 e_0/{\it d} U^2$ 
with $\ln(1/L)$ at $U=2.21t$ (empty squares), $2.225t$ (filled circles), $2.27t$ (empty 
circles).  The solid fitting -0.00997+0.009$\ln(1/L)$ is the least square fitting for 
the $U=2.225t$ case.}
\label{deri}
\end{figure}

At a finite chain length $L$, the ground-state energy per site $e_0(L)$ is also a function 
of $U$. Here $e_0(L)$ contains only the errors from the DMRG truncation.  We can examine 
the derivatives of $e_0(L)$ with respect to $U$ and analyze their chain length dependence.
In Fig. \ref{deri}(a), we show the  second derivative of $e_0(L)$ with $U$ for $L$=500, 
600, 800, and 1000.  At each chain length, there is a minimum near $U_c$ in the $U$ 
dependence of the second derivative.  When the chain length increases , the position of the
minimum moves towards larger $U$  and approaches 
$U_c$ which is the critical point in the thermodynamic limit; meanwhile, the
shape of the minimum becomes sharper.  Fig. \ref{deri}(b) shows the chain length dependence 
of the second derivative.  It is clear that for $U=2.225t \sim U_c$, the second derivative
diverges logarithmically with the chain length, but for other values of $U$ the second 
derivative does not diverge. These observations suggest that the phase transition at $U_c$ is 
of the second order.  No singularity is found in the first derivative or the second
derivative near the critical point $U_s$.  This means that the transition at $U_s$ is higher
than second order.  These results are consistent with those reported by 
Fabrizio {\it et al.}\cite{Fabrizio}.

\section{TMRG study of spin susceptibility and specific heat}
\label{secdmrg}

\begin{figure}[h!]
\includegraphics[width=8.0cm,angle=0]{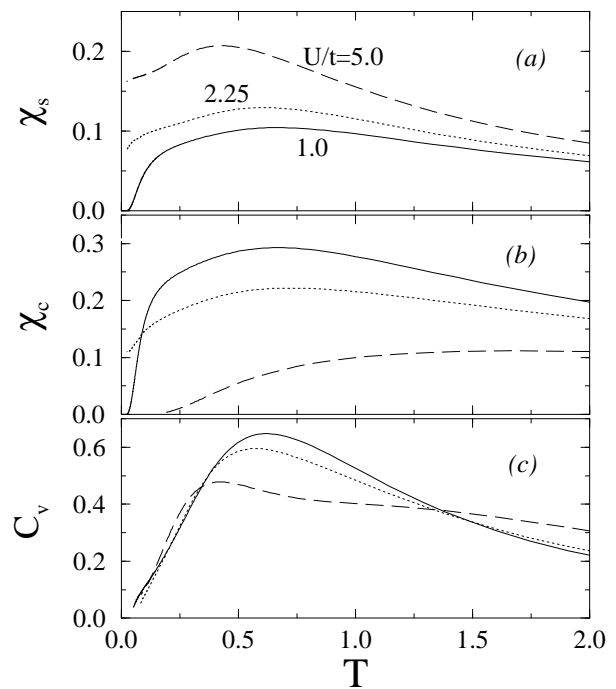}
\caption{Temperature dependences of (a) the spin susceptibility $\chi_s$, (b) the charge 
susceptibility $\chi_c$, and (c) the specific heat $C_v$. ($t$ is set to 1).}
\label{tmrg}
\end{figure}

To gain more insight into the physics of the BI-to-MI transition, we have studied the 
thermodynamic properties of the model using the TMRG method \cite{Bursill,Wang,Shibata} 
which is implemented in the thermodynamic limit and can evaluate very accurately the
thermodynamic quantities at low temperature for quasi-1D systems.  In our calculations,
we kept 250 optimal states.  The calculated specific heat $C_v$, charge 
susceptibility $\chi _c$, and spin susceptibility $\chi _s$ for $ U/t= 1.0, 2.25, 5.0$ and 
$V/t=0.3$ as a function of temperature are shown in Fig. \ref{tmrg}. 

We find that $\chi _c$ decreases exponentially at low temperatures in both the BI and MI 
phases, while $\chi _s$ shows activated behavior only in the BI phase. In the MI phase, 
there are two broad peaks in $C_v$, probably due to the charge-spin separation. Near the 
critical point, U=2.25t, since both the charge and spin energy gaps are very small, the 
exponential decays in both $\chi_c$ and $\chi_s$ show up only at very low temperatures.  
These results support the conclusions of the DMRG calculations presented in previous sections.

\section{Phase Diagram}
\label{phasediag}

The overall $U$ dependence of the charge, spin  and exciton gaps shown 
in Fig. \ref{bdu}  give a lot of information  on the phase diagram
of Hamiltonian (\ref{Ham}).  The charge and spin gaps coincide in the BI phase.  
Above $U_s$, $\Delta _c$ increases with $U$ but $ \Delta _s$ remains zero. 
The exciton gap $\Delta_e$ collapses at $U_c < U_s$, and when $U > U_s$, the exciton gap 
should also be zero. However, from Fig. \ref{bdu} it is unclear whether the exciton 
excitations are gapful or gapless in the regime between the two critical points,
$U_c < U < U_s$. Even if the exciton excitations are gapful in this regime, the gap 
would be too small to detect numerically. 

\begin{figure}[h!]
\includegraphics[width=8.0cm,angle=0]{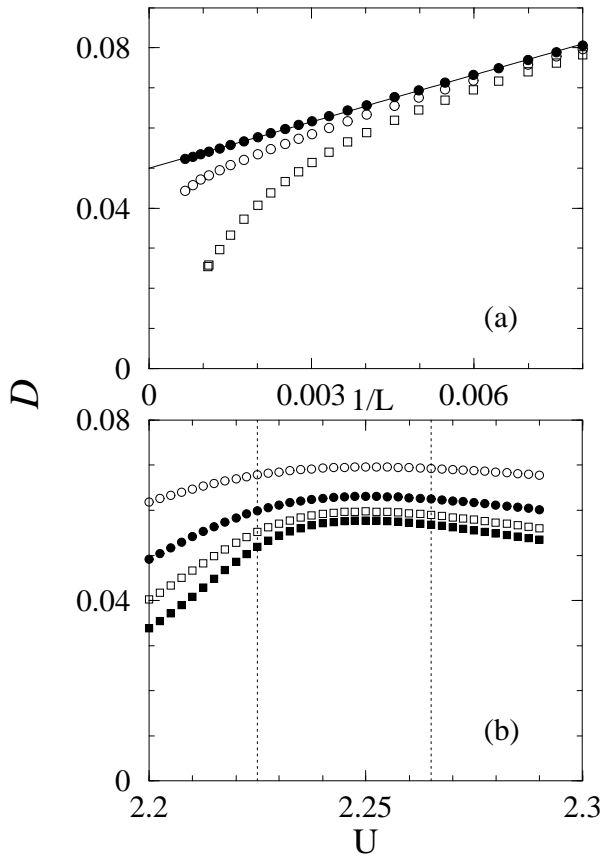}
\caption{(a) Chain length dependence of the dimerization order parameter ${\cal D}$ for 
$U=2.21t$ (empty squares), $U=2.25t$ (filled circles) and $U=2.29t$ (empty circles). 
The straight fitting line is $0.0499+3.886/L$.  (b) ${\cal D}(L)$ in the critical regime 
for $L$=200 (empty circles), 300 (filled circles), 400 (empty squares), and 500 (filled 
squares).  The dotted lines indicate the two critical points.}
\label{dimerfig}
\end{figure}

When the exciton gap collapses, the excitons can condense into the ground state \cite{aff}.  
In this case, the system is expected to be dimerized\cite{Fabrizio}.  Here we evaluate
the dimerization order parameter
\begin{equation}
{\cal D} = \frac{1}{L} \sum_{i \sigma} (-1)^i (c^\dagger_{i\sigma}
c_{i+1\sigma} + h.c.).
\label{dimereq}
\end{equation}
Figure \ref{dimerfig}(a) shows the chain length dependence of the dimerization operator 
for different $U$. It is clear that for $U=2.21t$, when $L \rightarrow \infty$, 
${\cal D}$ approaches zero. At $U=2.29t$, ${\cal D}$ just starts to fall at the largest 
chain length we studied; it is expected that it will approach zero as the chain length is
long enough.  For $U=2.25t$, which is between the two critical points, it seems that 
${\cal D}$ will diverge to a nonzero constant.  For a large range of chain lengths, the 
results can be well fitted by a straight line shown in Fig. \ref{dimerfig} (a).
The dependence of the finite chain dimerization ${\cal D}(L)$ on $U$ is shown in 
Fig. \ref{dimerfig}(b) for $L$=200, 300, 400, and 500.  These results indicate that in
the thermodynamic limit, the ground states are dimerized when $U_c < U < U_s$.

The dimerization of the ground state for $U_c < U < U_s$ suggests that the exciton 
excitations are gapful in this region. So the physical picture on the exciton excitation
is emerging: the exciton gap formed in the BI phase collapses at the critical point $U_c$;
with further increasing $U$, the (small) exciton gap will first increase, reach a maximum 
and then decrease and collapse again at $U_s$; at $U > U_s$, the exciton excitations
remain critical.

\begin{figure}[h!]
\includegraphics[width=8.0cm,angle=0]{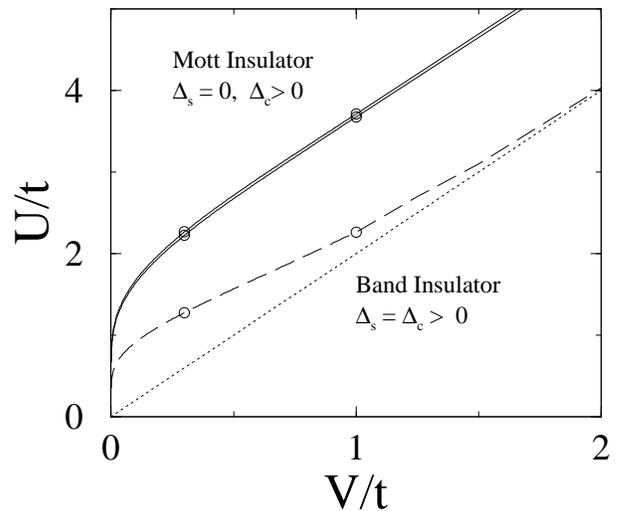}
\caption{The ground-state phase diagram for Hamiltonian (\ref{Ham}).  The empty circles 
denote the DMRG results. The curves of $U_s$ and $U_c$ are shown with solid lines that are 
very close to each other. The curve of $U_e$ is shown by the dashed line which does not 
indicate a phase transition line, but rather denotes a serial of special points. The dotted 
line is $U=2V$ which is the limit for $V \to \infty$.}
\label{phase}
\end{figure}

When $U\ll V$, first order perturbation leads to
\begin{equation}
\Delta _s(U)=\Delta _c(U)\approx V-cU,  
\label{deltas}
\end{equation}
where $c=V\int_0^\pi d\ k (\sqrt{2}\pi {\varepsilon }_k)^{-1} $ is a constant determined 
by the single-particle energy dispersion $ \varepsilon _k=\sqrt{V^2+4t^2\cos ^2k}$. 
Since $\Delta _c(U)$ drops almost linearly with $U/t$ in the BI phase, we can estimate the 
value of $U_s$ from Eq. (\ref{deltas}) as $U_s\approx V/c$. In the limit $V \to 0$, we have 
\begin{equation}
V/t \approx c_1e^{-c_2t/U_s},   ~~~~or ~~~~~
U_s/t \approx -{\frac{c_2}{\log (V/c_1 t)}}\ ,
\label{smallv} 
\end{equation}
where $c_1$ and $c_2$ are two constants of order one.  Figure \ref{phase} shows the 
ground-state phase diagram for Hamiltonian (\ref {Ham}). The curve for $U_s$ and $V <1.0t$ 
is obtained from Eq. (\ref{smallv}).  The parameters $c_1$ and $c_2$ are fixed 
by the two $U_s$ values for $V=0.3t$ and $V=1.0t$. When $V\rightarrow 0$, $U_s$ goes to 
zero but the ratio $U_s/V$ diverges.  In the limit $U/t\rightarrow \infty $, $U_s$ is very
close to $2V$.  The difference between $U_s$ and $2V$ is of order $t$: $U_s-2V\sim t$.

\section{Summary}
\label{sumdiscus}

We have carried out systematic studies using the DMRG and TMRG methods to examine
the critical behavior of a one-dimensional Hubbard model with an alternating site potential
in the transition from band insulator to Mott insulator.  Based on extensive numerical 
calculations and analytic analysis, we have clarified several important issues raised in 
recent works and have established the ground-state phase diagram.  
We have identified two critical points, $U_c$ and $U_s$, that separate the BI and MI phases.
When $U>U_s$, the system is in the MI phase where the charge excitations are massive but
the spin excitations are critical.  When $U<U_c$, the system behaves like a classic band
insulator: the charge and spin excitation gaps coincide and a charge-neutral spin-singlet
exciton band forms below the band gap when $U$ exceeds a special point $U_e$.  Between the
two critical points, excitons condense and the ground state is dimerized.  These results are 
consistent with the conclusions of a recent field theoretical study of the same model.  
The present work provides a detailed account for the critical behavior in the BI-to-MI 
transition in one dimension for correlated electron systems and establishes a good
understanding for its ground-state phase diagram.

\begin{acknowledgments}
We thank Y. L. Liu, R. Noack , and M. Fabrizio for helpful discussions and M. Tsuchiizu for 
bringing to our attention Refs. \onlinecite{tsu99} and \onlinecite{tsuc}.  This work was 
supported in part by the Department of Energy at the University of Nevada, Las Vegas, the NSF 
of China and the Special Funds for Major State Basic research Projects of China.
\end{acknowledgments}

\end{document}